# Resonant quenching of plasmon energy dissipation in a metal film with nonlocal dielectric response

**Vladimir S. Grigoryan**

*5227 Grovemont Drive, Elkridge, MD 21075, USA*
*vsgrigory@gmail.com*

**Evanescent waves in a metal thin film with nonlocality are found to propagate in normal direction to film surface with quenched (to zero) energy dissipation associated with intra-band electron transitions when wave numbers satisfy a resonant condition. It is shown that resonant quenching of energy dissipation (RQED) effect occurs in metal films with thicknesses of less or larger than, but still on the order of, the nonlocality scale length. RQED ceases to exist in metal films whose thickness exceeds a cutoff length or in metal films with local dielectric permittivity. Resonant quenching of energy dissipation is caused by destructive interference of partial contributions to electric displacement field, spatially dispersed over thin film thickness. It is demonstrated that RQED effect can be used for designing a new type of plasmonic waveguides, such as a slit waveguide representing a metal film with a narrow slit filled with a dielectric, to achieve near zero propagation losses for plasmonic modes with few nanometer scale confinement.**

## Significance

Plasmonic modes in existing plasmon waveguides with truly nanometer scale mode confinement, propagating in metals as evanescent waves, suffer from extremely high propagation losses associated with intra-band electron transitions, hindering their use for intra-chip interconnects and large scale plasmonic integrated circuits. In this work I show that there exists a special type of plasmonic mode with resonant wave numbers capable of evanescent propagation in a nonlocal metal thin film in normal direction to film surface without energy dissipation associated with the intra band electron transitions. Such waveguides can become a building block for a disruptive technology in truly nanometer-scale ultra-high quality factor plasmonic resonator cavities and electrically pumped ultra-low threshold and high slope-efficiency plasmonic lasers.

## I. Introduction

Over the past decade there has been an explosive interest in various metal-dielectric waveguide structures, such as the metal-insulator-metal plasmon waveguides, which enable deep sub-wavelength confinement of optical mode on the scale of tens of nanometers and below owing to negative real part of the metal dielectric permittivity [1] Realization of such waveguides in a practical way would have a potential of not only revolutionizing the area of intra-chip interconnects but also creating a fundamental building block for truly nanometer scale very large scale plasmonic integrated circuits. However, as the local dielectric response theory shows [1], deep sub-wavelength confinement of plasmons in the metal-dielectric waveguides comes at a steep price of extremely high losses associated with a significant imaginary part of the metal dielectric permittivity. Moreover, as the confinement becomes tighter, the material loss increases further, scaling near proportionally with the degree of confinement. Use of structures, such as the long-range plasmon waveguides [2] or other alternatives with reduced mode localization in the metal [3, 4] is hardly an option here as the modes in these structures spread widely in the dielectric and, as a consequence, confinement turns to be far from the desired scale. A straightforward way of reducing the loss is by adding materials with very high optical gain [5-7] makes the structures too complex to be practical, excessively noisy [8], and prone to heat dissipation problems. An alternative approach of creating a band pass transparency window by synthesizing a new type of metallo-crystal material with increased inter-atomic distance in the metal lattice by a factor of two relative to the inter-atomic distances naturally occurring in metals [9] appears to be too arduous and hypothetical undertaking to become practical as it is not clear how to do that and whether it is realizable by and large. Thus, in order to achieve a near-lossless plasmon propagation with deep sub-wavelength confinement, one has to resolve two fundamental problems. The first one is to provide a physical insight into a practically realizable mechanism capable of quenching energy dissipation in an existing metal and establishing limits for metal parameters and dimensions needed for that. The second one is to actually design a metal-dielectric waveguiding structure in which the metal operates in the dissipation quenching regime.

The current paper primarily addresses the first problem. It shows that the nonlocality in a metal nano-scale thin film results in spatial resonances of the metal dielectric response. Exploiting these spatial resonances, one can attain a zero imaginary part of the metal effective dielectric permittivity (with its negative real part) and thereby completely quench the energy dissipation associated with intra-band electron transitions for an evanescent plasmon wave with resonant wave numbers. The paper also demonstrates an example of solving the second problem. It presents a design for a slit plasmonic waveguide with deep sub-wavelength confinement and quenched to zero energy dissipation associated with the intra-band electron transitions.

The paper is organized as follows. In Section II, I show that RQED is qualitatively compliant with basic quantum mechanical principles. In Section III (along with Appendix A), a key dynamic equation for a plasmonic wave in a metal slab with nonlocal dielectric permittivity is derived within the macroscopic model. In Section IV, a dispersion relation and dielectric response in a nonlocal metal film are presented. Section V oulines a detailed description of evanescent waves with RQED in an Au film. In Section VI (along with Appendix B), I demonstrate that RQED effect can be used for designing a new type of plasmonic waveguides, a slit waveguide comprising a metal film with a narrow slit filled with a dielectric, to achieve a near lossless propagation for a plasmon mode with few nanometer scale confinement. Finally, the findings are summarized in Section VII.

## II. Feasibility for resonant quenching of plasmon energy dissipation on microscopic level

On a microscopic level, energy dissipation of a plasmon wave in a metal occurs when a conduction electron absorbs the energy of a photon and then loses its energy due to the scattering by phonons (both bulk and surface phonons). In this paper I will neglect absorption by defects. Therefore, one can argue that if one could prevent the electrons from absorbing the photons, the plasmon wave energy dissipation would be quenched. That is, plasmon energy would not dissipate. The key mechanism of the energy dissipation quenching can be understood within a quantum model for free electron gas.

Suppose a photon with a wave vector $\mathbf{k} = (k_x, 0, k_z)$ propagates is a metal film of thickness $d$ as shown in Fig. 1. When the photon is absorbed by an electron, the electron momentum changes from $\mathbf{p}$ (before the electro-photon interaction) to $\mathbf{p}'$ (after the electro-photon interaction). The interaction takes place only when the total momentum is conserved [10], i. e.

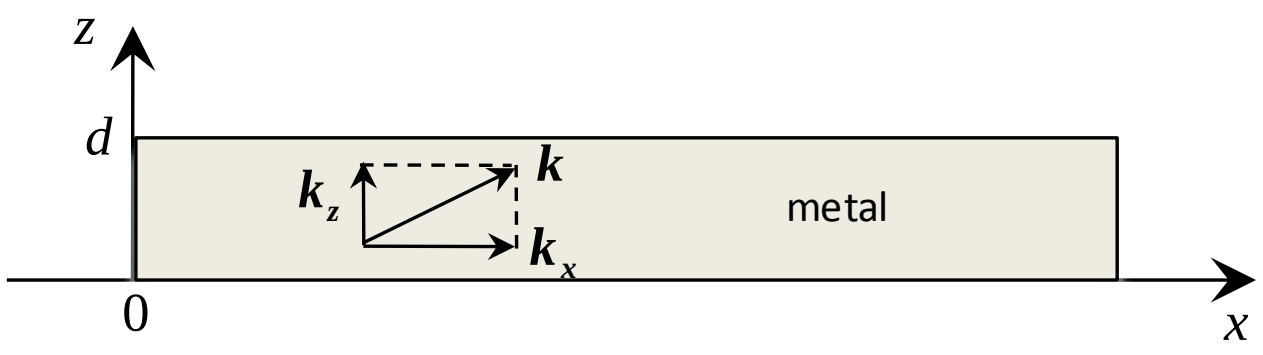

Fig. 1. Schematic diagram of a metal film with photon wave vector layout.

$$\mathbf{p}' = \mathbf{p} + \hbar\mathbf{k}, \tag{1}$$

assuming that both $\mathbf{p}$ and $\mathbf{p}'$ lie in the first Brillouin zone. On the other hand, the $z$-components of the electron momentum $p_z$ in the metal film are quantized obeying the Bohr-Sommerfeld quantization rule [11, 12]

$$\frac{2}{\hbar} p_z d + \Phi(E) = 2\pi n, \tag{2}$$

where $\Phi(E)$ is the cumulative phase shift of the electron wave function due to reflections from upper and lower boundaries of the film depending on the perpendicular energy $E = p_z^2 / 2m_e$ associated with the $z$-component of the electron momentum $p_z$ with $m_e$ being the electron mass, $n$ is an arbitrary integer number. When the perpendicular energy $E$ for the free electrons inside the thin film varies over its entire range from zero to the vacuum energy $E_V$, the cumulative phase shift $\Phi(E)$ varies from $-\pi$ radians at $E = 0$ to zero radians at $E = E_V$ [13, 14]. Assuming that both $\mathbf{p}$ and $\mathbf{p}'$ obey Bohr-Sommerfeld quantization rule, we have

$$\frac{2}{\hbar} p_z' d + \Phi(E) = 2\pi n' \tag{3}$$

in addition to Eq. (2), where $n'$ is another arbitrary integer number. Subtracting Eq. (2) from Eq. (3) and using that $p_z' - p_z = \hbar k_z$, following Eq. (1), one has

$$k_z d = \pi(n' - n) - \delta\Phi/2, \tag{4}$$

where $\delta\Phi = \Phi(E') - \Phi(E)$. Now, if the photon wave vector is such that $k_z = 2\pi m/d$, where $m$ is an arbitrary integer number, then it must be, from Eq. (4),

$$2\pi m = \pi(n' - n) - \delta\Phi/2. \tag{5}$$

However, Eq. (5) can never be satisfied because $-\pi < \delta\Phi/2 < \pi$ for any nontrivial pair of energies $E$ and $E'$ as far as $-\pi < \Phi(E) < 0$ for any $E$. Consequently, given the Bohr-Sommerfeld quantization rule, the electrons cannot absorb photons with wave vector *z*-components $k_z = 2\pi m/d$ (where $m = 0, \pm 1, \pm 2, ...$) without violating the momentum conservation law. Or, in a more general sense, without violating at least one of the two relations, the momentum conservation law or the Bohr-Sommerfeld quantization rule. Therefore, photons with resonant *z*-components $k_z = 2\pi m/d$ (of their wave vector) propagating in a metal thin film of thickness *d*, will not dissipate their energy. It would have been utterly deludable to argue that such a result derived merely from the momentum conservation law and Bohr-Sommerfeld quantization rule could have been completely overlooked in the literature. It was not. However, the key argument was that the microscopic model comprising the momentum conservation law and Bohr-Sommerfeld quantization rule was too simplistic and not complete. It did not take into account the electron collisions with phonons, which would cause energy dissipation. Besides, the microscopic model also did not account for the hydrodynamic evolution of the electron gas induced by the optical field. The perception was that once these mechanisms had been accounted for, photons would always dissipate. The perception was, indeed, confirmed by the macroscopic model with a local dielectric permittivity where both the electron collisions with phonons and hydrodynamic evolution of the electron gas were accounted for (on the macroscopic level). In Section V, I show that the nonlocality changes the picture qualitatively. It turns out that within the macroscopic model with nonlocal dielectric permittivity there exist some resonant wavenumbers at which the electromagnetic waves can propagate in metal thin films with no energy dissipation. It is demonstrated in Section V that those resonant wavenumbers are not equidistantly spaced for a finite collision frequency, as was predicted by the Bohr-Sommerfeld quantization rule – based microscopic model. However, when the collision effect becomes negligible, the resonant wave numbers become equidistantly spaced, with 2π/d period, which is consistent with the $k_z = 2\pi m/d$ formula predicted the Bohr-Sommerfeld quantization rule – based microscopic model.

Furthermore, on the macroscopic level, resonant quenching of energy dissipation means that the imaginary part of the dielectric response becomes zero at the resonant wave numbers $k_z = 2\pi m/d$ and non-zero elsewhere, resulting in an oscillatory behavior of the imaginary part of the dielectric response with film thickness *d* at a given wave number $k_z$. Drawing an analogy between the spatial dispersion (spatial nonlocality) and chromatic dispersion (temporal nonlocality), these oscillations turn to be similar to Rabi oscillations of the quadrature component of atom dipole momentum in the time domain [15] for coherent interactions of light with resonant atoms. Akin to our case of RQED in metal films with spatial nonlocality, in the coherent interactions with resonant atoms, optical power dissipation rate also turns to zero at a time at which the quadrature component of the atom dipole momentum becomes zero.

Next, as was pointed out above, a simple quantum model based on the Bohr-Sommerfeld quantization rule, described above does not take into account an impact of electron collisions on the resonant wave numbers as well as it does not account for hydrodynamic evolution of the electron gas induced by optical field. Analyzing these effects and their impact on RQED within a full quantum model appears to be a hardly efficient approach. On the other hand, these effects can be implicitly accounted for in the classical macroscopic model based on Maxwell's equations with nonlocal dielectric response. That is why, a generic macroscopic model for a metal with nonlocality is applied below to analyze the RQED effect and determine its limits.

## III. Basic equations

Within the classical macroscopic model, interaction of light with matter is described by Maxwell's equations,

$$\nabla \times \mathbf{E}(\mathbf{r},\omega) = -i\frac{\omega}{c}\mathbf{B}(\mathbf{r},\omega), \quad \nabla \times \mathbf{H}(\mathbf{r},\omega) = i\frac{\omega}{c}\mathbf{D}(\mathbf{r},\omega), \tag{6a}$$

$$\nabla \cdot \mathbf{D}(\mathbf{r},\omega) = 0, \quad \nabla \cdot \mathbf{B}(\mathbf{r},\omega) = 0, \tag{6b}$$

where $\mathbf{E}(\mathbf{r},\omega)$ and $\mathbf{H}(\mathbf{r},\omega)$ are the electric and magnetic field vectors respectively, $\mathbf{B}(\mathbf{r},\omega) = \mu\mathbf{H}(\mathbf{r},\omega)$ is the magnetic induction vector with $\mu$ being the magnetic permeability constant,

$$\mathbf{D}(\mathbf{r},\omega) = \int \varepsilon(\mathbf{r},\mathbf{r}',\omega)\mathbf{E}(\mathbf{r}-\mathbf{r}',\omega)d\,\mathbf{r}', \tag{7}$$

is the electric displacement vector in the medium with nonlocality (spatial dispersion) where $\varepsilon(\mathbf{r},\mathbf{r}',\omega)$ is a spatially dependent (nonlocal) and chromatically dispersive dielectric permittivity function. Note that all the vector fields in Eqs. (6) and (7) are $\omega$-frequency Fourier components of their respective fields in the time domain at the spatial radius vector $\mathbf{r}$. Only magnetically local media are of interest in this paper and, therefore, the nonlocality of magnetic induction is neglected. In a homogeneous medium $\varepsilon(\mathbf{r},\mathbf{r}',\omega) = \varepsilon(|\mathbf{r}'|,\omega)$ and therefore, in the spatial spectrum domain ($\mathbf{k}$-space), Eq. (7) can be written as

$$\mathbf{D}(\mathbf{k},\omega) = \varepsilon(\mathbf{k},\omega)\mathbf{E}(\mathbf{k},\omega). \tag{8}$$

The dielectric permittivity function $\varepsilon(\mathbf{k},\omega)$ for a metal like Ag or Au is defined as

$$\varepsilon(\mathbf{k},\omega) = \varepsilon_\infty + \varepsilon_{\mathrm{inter}}(\omega) + \varepsilon_{\mathrm{intra}}(\mathbf{k},\omega), \tag{9}$$

where $\varepsilon_\infty$ is the $\varepsilon(\mathbf{k},\omega)$ value at $\omega \to \infty$, $\varepsilon_{\text{inter}}(\omega)$ is the contribution from *d*-band to conduction-band (*sp*-band) electronic transitions, and $\varepsilon_{\text{intra}}(\mathbf{k},\omega)$ is due to the excitations of conduction electrons [16, 17]. In the short wavelength infrared spectral region of our interest (frequencies lower than 1.23984 eV), the contribution of the inter-band transitions $\varepsilon_{\text{inter}}(\omega)$ turns to be negligible relative to the $\varepsilon_{\text{intra}}(\mathbf{k},\omega)$ as these frequencies are substantially lower than the onset frequencies for the inter-band transitions (3.9 eV for Ag [18-20] and 2.3 eV for Au [18, 20]). As to the intra-band contribution $\varepsilon_{\textbf{intra}}(\mathbf{k},\omega)$, its form is derived based on the density functional theory formalism [21],

$$\varepsilon_{\text{intra}}(\mathbf{k},\omega) = -\frac{\omega_{\text{p}}^2}{\omega^2 + i\gamma\omega - \beta^2\mathbf{k}^2}, \tag{10}$$

where $\omega_{\text{p}}$ is the plasma frequency, $\gamma$ is the collision frequency, and $\beta^2$ is the nonlocality parameter which, for free electron gas in a thin film – like system, is $\beta^2 = v_{\text{F}}^2$ [16, 17], where $v_{\text{F}}$ is the Fermi velocity. It is shown in Appendix A that the dynamic equations describing a transverse magnetic (TM) wave $\boldsymbol{\mathcal{E}} = (\mathcal{E}_x, 0, \mathcal{E}_z)$ propagating in a metal slab (or film) lying parallel to the (*x*,*y*) plane are as follows,

$$\frac{\partial D_z}{\partial z} = \frac{\omega}{2\pi c}\int_{-d/2}^{d/2} \varepsilon(\zeta,\omega) f(z-\zeta,\omega)\, d\zeta, \tag{11a}$$

$$\frac{\partial f}{\partial z} = \frac{2\pi c}{\omega}\left(k_x^2 \mathcal{E}_z - \mu\frac{\omega^2}{c^2} D_z\right), \tag{11b}$$

where $D_z = \hat{\varepsilon}\mathcal{E}_z = \int_{-d/2}^{d/2} \varepsilon(\zeta,\omega)\mathcal{E}_z(z-\zeta,\omega)\, d\zeta$, $d$ is the slab (or film) thickness, $\mathcal{E}_z(z,\omega)$ is the amplitude of the *z*-component of the electric field vector of the TM wave having a fixed *x*-component $k_x$ of the wave vector $\mathbf{k}$, and the nonlocal dielectric function $\varepsilon(\zeta,\omega)$ is defined as,

$$\varepsilon(\zeta,\omega) = \varepsilon_\infty \delta(\zeta) + \frac{i\omega_p^2}{2a}\exp(ia|\zeta|), \tag{12}$$

where

$$a^2 = \frac{1}{\beta^2}(\omega^2 + i\gamma\omega) - k_x^2. \tag{13}$$

## IV. Dispersion relation and dielectric response in a metal film with nonlocality

We are looking for a solution of Eqs. (11) in the form,

$$D_z(z) = C_1 \exp(ibz), \quad f(z) = C_2 \exp(ibz), \quad \mathcal{E}_z(z) = C_3 \exp(ibz), \tag{14}$$

with $C_{1,2,3} = C_{1,2,3}(\omega)$ being some constants independent of *z*. Substituting Eqs. (14) into Eqs. (11) we derive the following set of algebraic equations for $C_{1,2,3}$,

$$\begin{aligned} &-ibC_1 + \frac{\omega\varepsilon_B}{2\pi c}C_2 = 0, \\ &-ibC_2 + \frac{2\pi c}{\omega}\left(k_x^2 C_3 - \frac{\omega^2\mu}{c^2}C_1\right) = 0, \\ &C_1 - \varepsilon_B C_3 = 0, \end{aligned} \tag{15}$$

where

$$\varepsilon_B = \int_{-d/2}^{d/2} \varepsilon(\zeta,\omega)\exp(-ib\zeta)d\zeta . \tag{16}$$

Integration in Eq. (16), with $\varepsilon(\zeta,\omega)$ taken from Eq. (12), yields the following dependence of $\varepsilon_B$ on $d$ and $b$ (as well as on $k_x, \beta, \omega$ and $\gamma$ through $a$),

$$\varepsilon_B = \varepsilon_\infty + \frac{\omega_p^2}{2\beta^2 a}\left\{\frac{\exp[i(a+b)d/2]-1}{a+b} + \frac{\exp[i(a-b)d/2]-1}{a-b}\right\}. \tag{17}$$

A solution for the $\mathcal{E}_x$ component is

$$\mathcal{E}_x(z) = C_4 \exp(ibz), \tag{18}$$

where $C_4 = -C_1 b/\varepsilon_B k_x$ found by substituting $D_z$ from Eq. (14) into Eq. (15) (see derivation details in Appendix A). The equations (15) have a non-trivial solution only when

$$b^2 + k_x^2 = \frac{\omega^2}{c^2}\varepsilon_B \mu . \tag{19}$$

Thus, Eq. (19) represents a dispersion relation for a TM wave in a metal film with nonlocality. The form of this dispersion relation looks similar to that of a conventional isotropic, homogeneous medium with infinite dimensions having a dielectric permittivity $\varepsilon_B$ and a magnetic permeability $\mu$. However, unlike the later, the dielectric permittivity $\varepsilon_B$ for a metal film with nonlocality explicitly depends on film thickness $d$. In the limit of large thickness $d$, when $d$ is much larger than the nonlocality scale length $r_{\text{nl}} = [\text{Im}(a)]^{-1}$ (see Eqs. (A10) and (A11) in Appendix A, for more details) the dielectric response $\varepsilon_B$ tends to

$$\varepsilon_B\left(d \gg r_{\text{nl}}\right) = \varepsilon_\infty - \frac{\omega_p^2}{\beta^2\left(a^2 - b^2\right)} = \varepsilon_\infty - \frac{\omega_{\text{p}}^2}{\omega^2 + i\gamma\omega - \beta^2\left(b^2 + k_x^2\right)}, \tag{20}$$

and no longer depends on $d$ . Note that in this case the dielectric response $\varepsilon_B$ in Eq. (20) is fully consistent with the local dielectric permittivity formula as in the latter case the nonlocality scale length $r_{nl} = 0$ and the ratio $d/r_{\rm nl} = \infty$ for any finite $d$ .

## V. Spatial resonances and quenching of plasmon energy dissipation in a metal film

Equations (13), (17), and (18) represent a full set of equations to determine $k_x$ as a function of $b$ and *d*. Then, the dielectric response $\varepsilon_B$ can be also as a function of $b$ and *d* through Eqs. (17) and (13). To solve Eqs. (13), (17), and (18), we first set $b$ (for a given $\omega$ and *d*) and then find a solution for $k_x$. The following dielectric parameters for bulk Au are used in this paper: $v_{\rm F} = 1.39\times10^6$ m/s, $\omega_p = 8.812$ eV [16], whereas the bulk Au collision frequency $\gamma_0 = 0.0749$ was slightly adjusted relative to Ref. [16] and $\varepsilon_\infty$ was set to 5.189 to extrapolate the data on the real and imaginary parts of the Au dielectric permittivity [16] at the extended frequency $\omega =$ 0.799898 eV (1550 nm wavelength). The collision frequency increase due to the conduction electron – surface interface scattering was taken into account by approximating a film of thickness *d* by a cylinder with a large radius (much larger than its height) and height equal to *d* using the effective collision frequency $\gamma = \gamma_0 + 0.1 v_{\rm F}/(2d)$ [22, 23]. Figure 2(a) shows the real and imaginary parts of $\varepsilon_B$ depending on the film thickness *d* at a fixed wave number *b* of 2π/20nm and optical wavelength of 1550 nm in an Au film. One can see that there are some resonant thicknesses *d* (marked with dots) at which $\mathrm{Im}(\varepsilon_B)$, responsible for the plasmon energy dissipation, reaches zero. Similar resonances occur also in the wave number space for the resonant wave numbers *b* at fixed *d* as will be shown below in Fig. 3. Thus, at these resonances, the energy dissipation associated with the intra-band electron transitions is quenched to zero. Resonant quenching of energy dissipation can be interpreted as a destructive interference of partial contributions to the electric displacement field, spatially dispersed over the film thickness, in a nonlocal metal film. Areas with negative $\mathrm{Im}(\varepsilon_B)$ (marked with grey color in Fig. 2) designate spatial gaps as there is no physical solution that can exist there in passive media with *b* = 2π/20nm. However, it does not mean that no other solution with different *b* for any *d* within these spatial gaps can exist, as shown below in Fig. 3(a). Note that when the impact of the collision frequency $\gamma$ becomes small, such that the ratio $d/r_{\rm nl}$ ≪ 1 (also achieved when *d* is small), the resonant wave numbers *b* become equidistantly spaced, with 2π/*d* spacing, which is consistent with the $k_z = 2\pi m/d$ formula predicted in Section II based on the simplified quantum model. This follows directly from Eq. (17), and also can be noticed in Fig. 2(a) as a trend of equalizing the spacing between adjacent resonant wave numbers *b* as *d* decreases. As film thickness *d* increases, the impact of the collision frequency $\gamma$ becomes stronger and the resonant wave numbers (dots) in Fig. 2(a) significantly deviate from the equidistant spacing formula. As was pointed out above in Section II, there is an inherent analogy between the spatial nonlocality (spatial dispersion) and temporal nonlocality (chromatic dispersion) occurring in coherent interactions of light with resonant atoms. Indeed, the oscillations of the real and imaginary parts of the dielectric constant $\varepsilon_B$, shown in Fig. 2(a), turn to be similar to the Rabi oscillations of the in-phase and quadrature components of the atomic dipole moment in the time domain [15]. Like the imaginary part of $\varepsilon_B$, the quadrature component of the atomic dipole moment is also responsible for the energy dissipation.

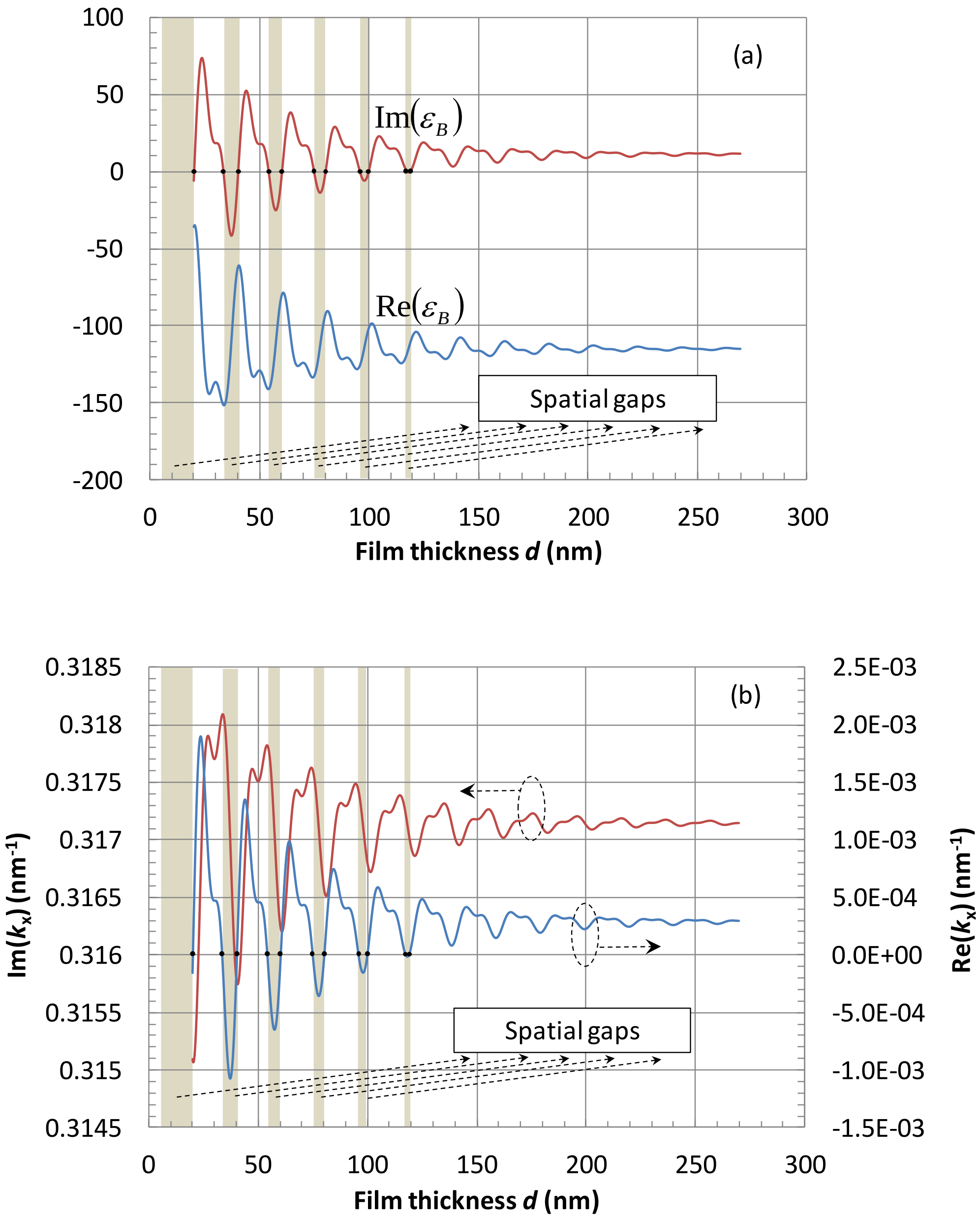

Fig. 2. Dependence of: (a) $\mathrm{Re}(\varepsilon_B)$, $\mathrm{Im}(\varepsilon_B)$ and (b) $\mathrm{Re}(k_x)$, $\mathrm{Im}(k_x)$ on thin film thickness *d* at fixed *b* of 2π/20nm in an Au film at optical wavelength of 1550 nm. Areas with $\mathrm{Im}(\varepsilon_B) < 0$ (spatial gaps) are marked with grey with dots designating the edges of the gaps.

However, while both signs of the quadrature components of the atomic dipole moment make physical sense, resulting simply in a redistribution of the optical power in the time domain, only a positive (or zero) imaginary part of $\varepsilon_B$ is physically meaningful in the nonlocal passive systems. Next, once *b* is set to 2π/20nm and $\varepsilon_B$ versus *d* is found in Fig. 2(a), $k_x$ wave number also becomes a function of *d*, as shown in Fig. 2(b). One can see that the resonant points (dots where $\mathrm{Im}(\varepsilon_B) = 0$) correspond to purely imaginary $k_x$ values. That is, the waves enabling the resonant quenching of energy dissipation are of the evanescent nature. Note in this regard that some resonant absorption minima (located near $b = 2\pi m/d$) were observed in Ref. [17] in numerical simulations for Au films. Although there was a trend for those absorption minima of becoming smaller as the energies moved farther below the 2.3 eV onset frequency for Au, the true resonances with zero energy dissipation were not detected there. Neither, the resonant zero energy dissipation had been observed earlier, both theoretically [24] and experimentally [25]. The reason for this was that the authors analyzed only the plane waves either at normal incidence [16] (zero $k_x$) or at angle incidence [24, 25] (non-zero but real $k_x$) to the film surface while there was no analysis done for the evanescent waves. As Fig. 2(b) shows, the $k_x$ wave numbers, corresponding to the resonant quenching of the energy dissipation, are always non-zero and imaginary. According to Eqs. (13), (17), and (18), the waves with zero $k_x$ or non-zero but real $k_x$ (plane waves) would have a non-zero energy dissipation (non-zero imaginary parts of $\varepsilon_B$). Both $\varepsilon_B$ and $k_x$ in Figs. 2, first, oscillate with *d* when *d* is less or larger than, but on the order of, the nonlocality scale length $r_{\mathrm{nl}} \approx 25$ nm, and then relax to their steady-state quantities. When *d* exceeds a cutoff length of about 120 nm, $\mathrm{Im}(\varepsilon_B)$ can no longer reach zero and, consequently, the resonant wave numbers no longer exist. As *d* increases further and becomes much larger than the nonlocality scale length $r_{\mathrm{nl}}$, $\varepsilon_B$ asymptotically approaches its steady-state quantity in accordance with Eq. (19) for bulk Au with $\mathrm{Im}(\varepsilon_B)$ of about 11.2 and $\mathrm{Re}(\varepsilon_B)$ of about –114.8 at *b* = 2π/20nm. In this case the steady-state $k_x$ of $k_x = -2.96\times10^{-4}\mathrm{nm}^{-1} \; - i0.317\mathrm{nm}^{-1}$ simply follows from Eq. (19). Note that the observation made above that the resonant quenching of the energy dissipation may exist only when the film thickness is less or larger than, but still on the order of, the nonlocality scale length, remains valid for any metal film.

Figure 3(a) shows dependence of the dielectric response $\varepsilon_B$ on *b* for an Au film with fixed *d* =20 nm. One can see that there is a wide wave number gap in this case spreading over 0.233 $\mathrm{nm}^{-1}$ < *b* < 1.02 $\mathrm{nm}^{-1}$, where no physical solutions exist in a passive medium. On the other hand, when *b* < 0.233 $\mathrm{nm}^{-1}$ or *b* > 1.02 $\mathrm{nm}^{-1}$, the imaginary part of $\varepsilon_B$ is positive and can be substantially large indicating a significant energy dissipation. However, right at the edges of the gap, at $b = 0.233$ nm$^{-1}$ or $b = 1.02$ nm$^{-1}$ (resonant quantities of $b$ marked with dots), the imaginary part of $\varepsilon_B$ becomes zero and the energy dissipation is quenched. Comparing Fig. 3(a) with Fig. 2, one can see that for a thickness $d$ that ends up in a spatial gap in Fig. 2 for $b$ = 2π/20nm, a solution with different $b$ may exist. Indeed, in Fig. 2(a), $\mathrm{Im}(\varepsilon_B) = -6.05$ at $d = 20$ nm (belongs to the first spatial gap) for $b$ = 2π/20nm (0.314 nm$^{-1}$), whereas for the same $d = 20$ nm in Fig. 3(a) we have $\mathrm{Im}(\varepsilon_B) > 0$ at $b < 0.233$ nm$^{-1}$ or $b > 1.02$ nm$^{-1}$ and, hence, physical solutions in these areas of $b$ do exist. Figure 3(b) shows dependence of the real and imaginary parts of $k_x$ on $b$, representing a spatial dispersion relation for the same Au film with $d = 20$ nm. Note that at the resonant wave numbers $b = b_0 = 0.233$ nm$^{-1}$ or $b = b_0 = 1.02$ nm$^{-1}$, the real part of $k_x$ becomes zero whereas its imaginary part is positive with substantially large magnitude. Therefore, the resonant quenching of energy dissipation occurs for two evanescent waves with strong lateral decay rate along $x$-axis (about $k_x = k_{x0} = i0.236$ nm$^{-1}$ at $b = b_0 = 0.233$ nm$^{-1}$ and $k_x = k_{x0} = i1.02$ nm$^{-1}$ at $b = b_0 = 1.02$ nm$^{-1}$).

Finally, chromatic dispersion of the resonant wave numbers $b_0$ and $k_{x0}$ as well as the resonant dielectric response $\varepsilon_B = \varepsilon_{B0}$ for a 20 nm thick Au film is shown in Fig. 4(a). At 1550 nm wavelength, the $\varepsilon_{B0}$, $k_{x0}$ and $b_0$ data in Fig. 4(a) coincide with those for the left-hand side resonant $b = b_0 = 0.233$ nm$^{-1}$ in Fig. 3. Whereas, at 1550 nm wavelength, the $\varepsilon_{B0}$, $k_{x0}$ and $b_0$ data in Fig. 4(b) coincide with those for the right-hand side resonant $b = b_0 = 1.02$ nm$^{-1}$ in Fig. 3. One can see that in both cases the resonant dielectric response $\varepsilon_{B0}$ remains negative real while $k_{x0}$ stays positive imaginary over the wavelength range from 1530 nm to 1625 nm (optical C- and L-bands).

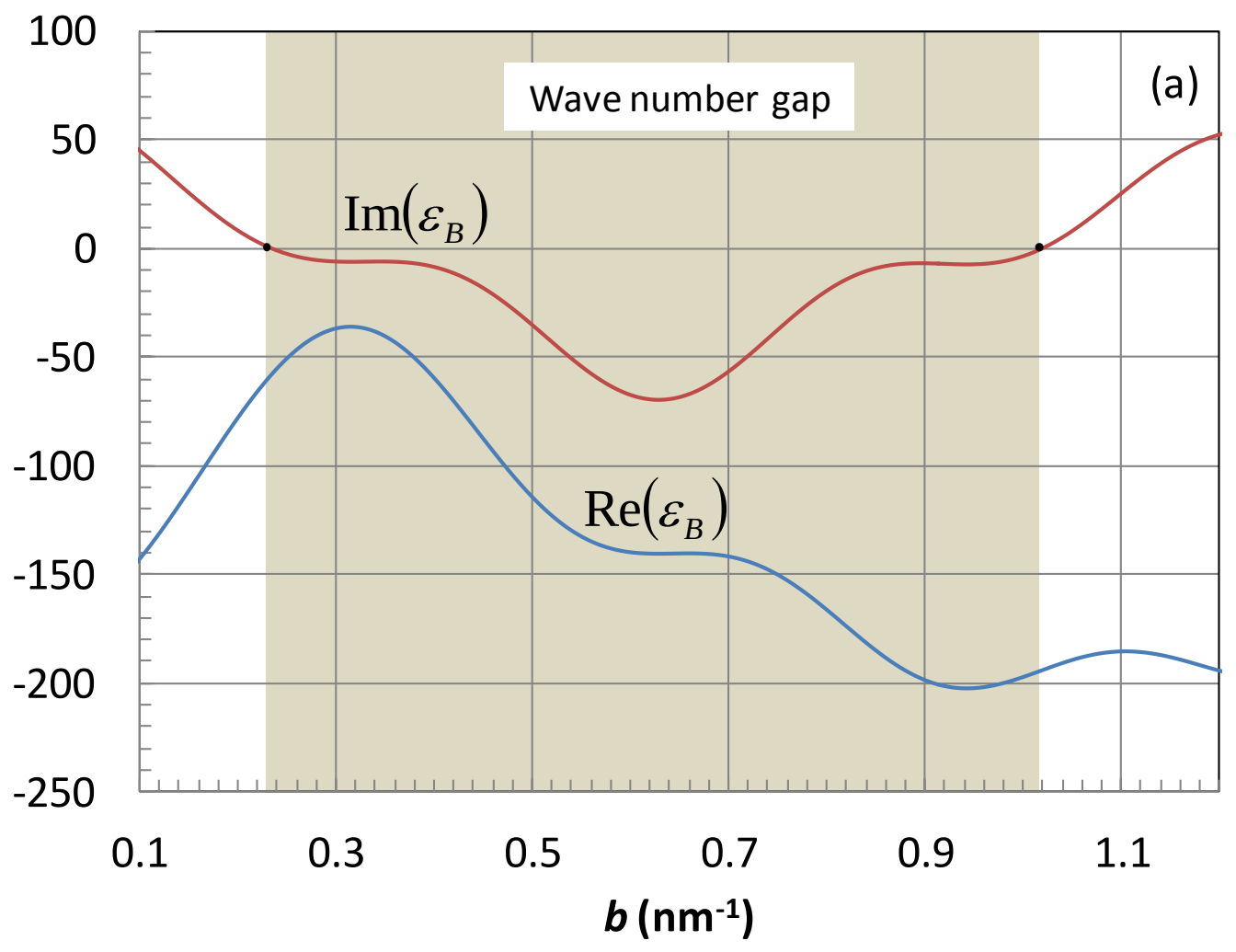

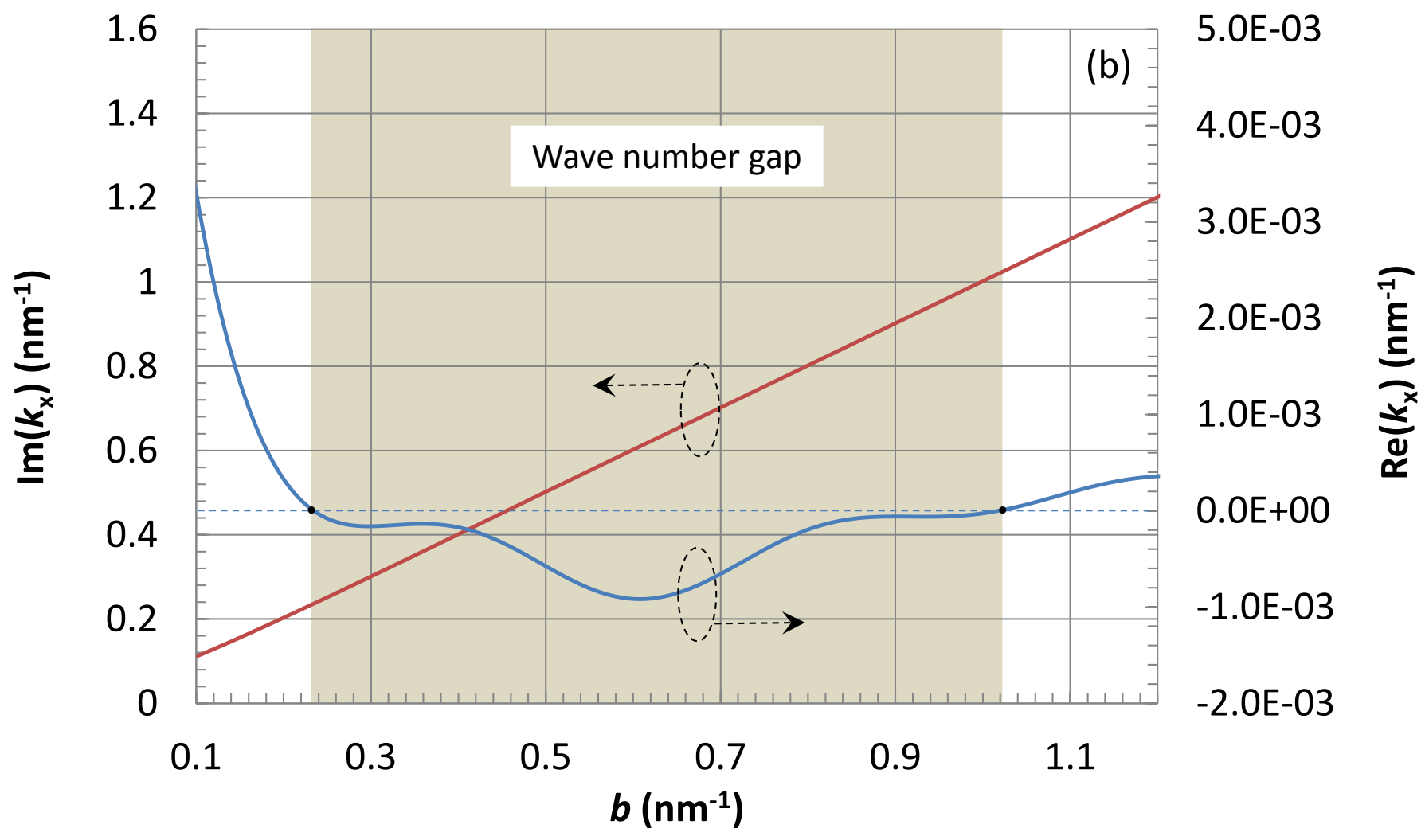

Fig. 3. Dependence of: (a) $\mathrm{Re}(\varepsilon_B)$, $\mathrm{Im}(\varepsilon_B)$ and (b) $\mathrm{Re}(k_x)$, $\mathrm{Im}(k_x)$ on wave number $b$ in an Au thin film with fixed thickness $d$ =20 nm at optical wavelength of 1550 nm. Area with $\mathrm{Im}(\varepsilon_B)<0$ (wave number gap) is marked with grey with dots designating the edges of the gap.

Note that the magnitude of $k_{x0}$ stays close to the magnitude of $b_0$ in accordance with Eq. (19), when each of them is significantly larger than $\omega\sqrt{|\varepsilon_B|\mu}/c$.

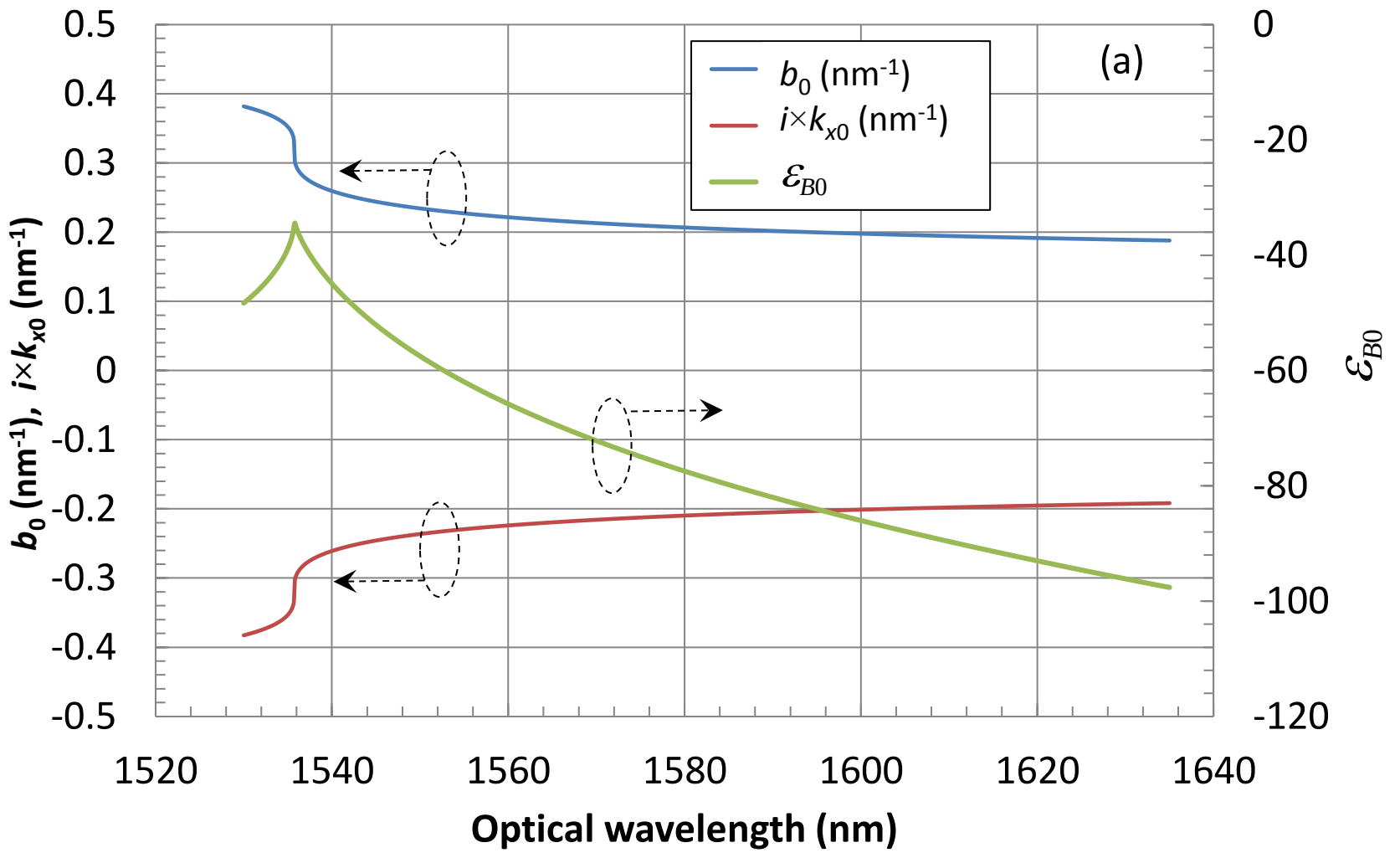

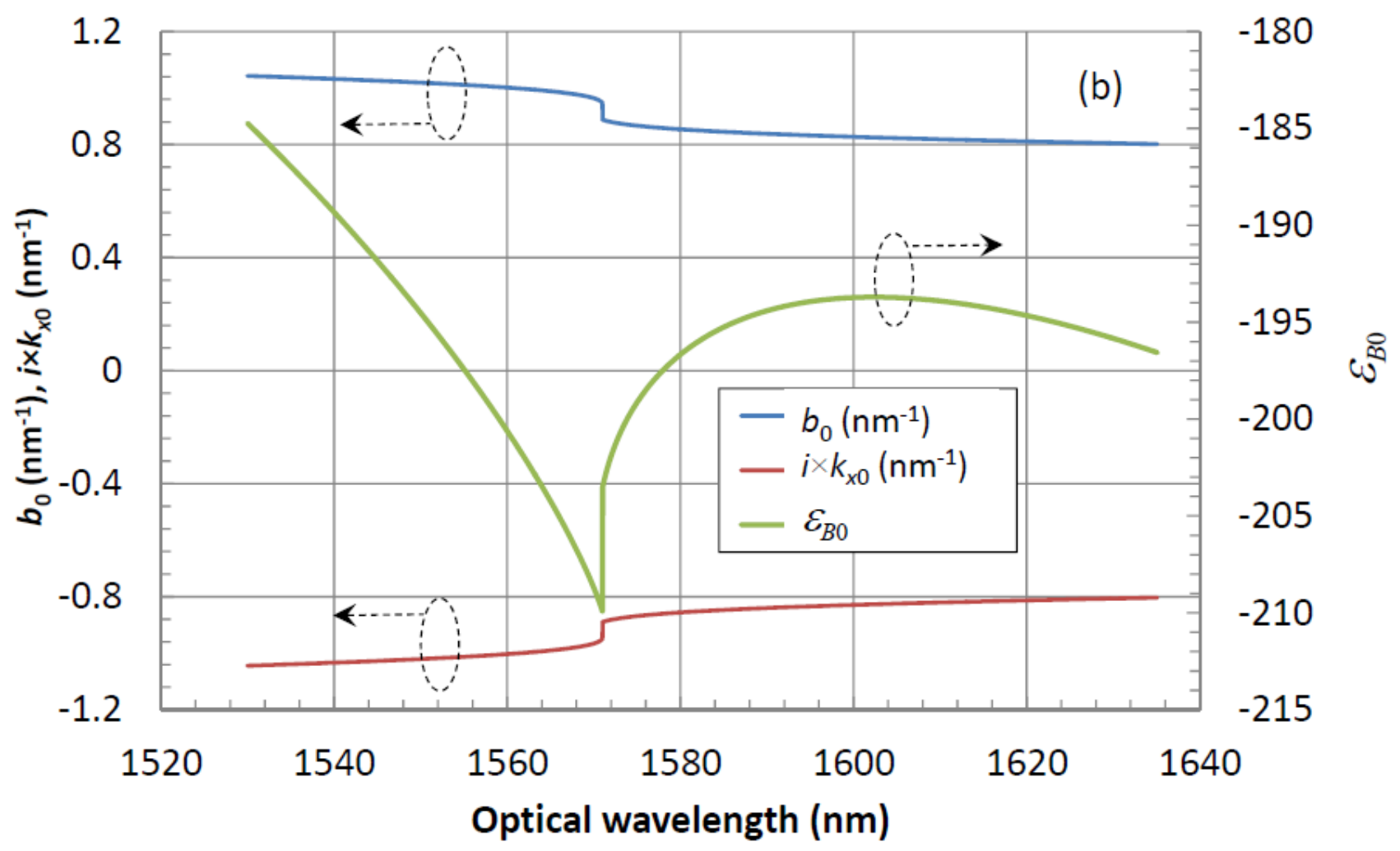

Fig. 4. Resonant wave numbers $b$ and $k_x$ as well as resonant dielectric response $\varepsilon_B$ versus optical wavelength for a 20 nm thick Au film (a) corresponding to the left-hand side resonant $b$ in Fig. 2 and (b) corresponding to the right-hand side resonant $b$ in Fig. 2.

## VI. Lossless propagation of plasmons in a slit waveguide

Ability of the resonant evanescent waves to propagate through a metal thin film in normal direction to film surface with quenched to zero energy dissipation (associated with intra band electron transitions) is a remarkable phenomenon that has a direct application to ultra-low loss plasmon waveguides with deep sub-wavelength confinement. Figure 5 shows a slit waveguide formed by two metal thin films and separated by a narrow slit filled with a dielectric.

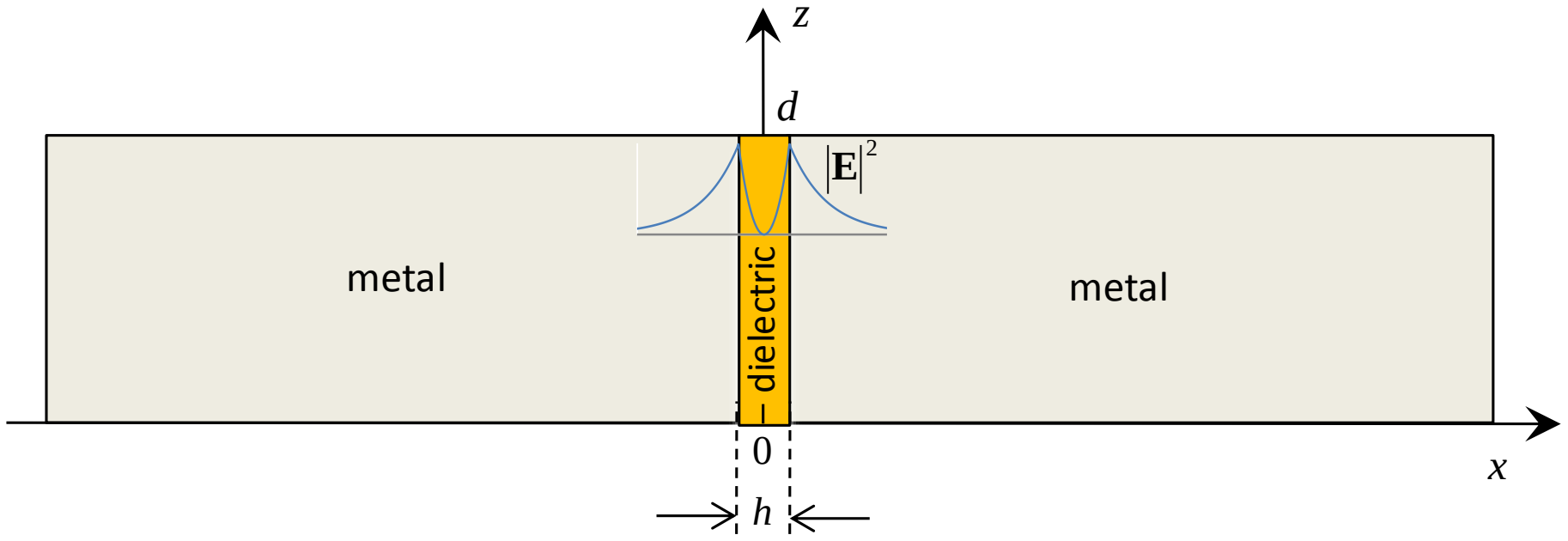

Fig. 5. Schematic diagram of a slit waveguide, comprising two metal thin films of thickness $d$ separated by a narrow slit of width $h$ filled a dielectric, for lossless propagation of a TM plasmon wave in $z$-direction.

As shown in Appendix B (see Eq. (B12)), the slit waveguide enables a TM plasmon wave propagating in $z$-direction with resonant evanescent tails spreading into the metal with no propagation losses (associated with the intra band electron transitions) if the width of the slit $h$ is such that

$$h = \frac{1}{ik_{x1}} \ln\left( \frac{\varepsilon_{B0} k_{x1} + \varepsilon_1 k_{x0}}{\varepsilon_{B0} k_{x1} - \varepsilon_1 k_{x0}} \right), \tag{21}$$

where $k_{x0}$ and $k_{x1}$ are the *x*-component resonant wave numbers in the metal film and dielectric slit respectively, $\varepsilon_1$ is the dielectric permittivity of the dielectric slit and $\varepsilon_{B0}$ is the resonant dielectric response (described above in detail in Section V). For a negative and real $\varepsilon_{B0}$, positive and real $\varepsilon_1$, positive imaginary $k_{x0}$ and imaginary $k_{x1}$ of both signs (for large enough $b_0^2 > \varepsilon_1 \mu \omega^2 / c^2$), Eq. (21) always has a non-trivial solution for *h*. Thus, for a given resonant evanescent wave (with the resonant wave numbers $k_{x0}$, $b_0$ and resonant dielectric response $\varepsilon_{B0}$), there exists a resonant width *h* for the dielectric slit, defined by Eq. (21), which supports a lossless plasmon wave with quenched energy dissipation. The field profile for such lossless plasmon wave is described in detail in Appendix B (see Eq. (B15)). Figure 6 shows wavelength dependence of the slit width *h* for the slit waveguide (outlined in Fig. 5) composed of *d* = 20 nm thick Au films capable of supporting resonant lossless plasmons with crystalline GaAs, InP, and Si slits. The inset in Fig. 6 illustrates one example for a mode field profile in the slit waveguide with a crystalline Si slit. In Fig. 6, the resonant wave numbers $k_{x0}$, $b_0$ and resonant dielectric response $\varepsilon_{B0}$ were taken from Fig. 4(a) with the chromatic dispersion of $\varepsilon_1$ for GaAs and Si taken from Ref. [26] and for InP from Ref. [27]. Since the nonlocality in the dielectric crystal slit is significantly smaller than that in a metal, the former has been neglected. Accounting for the dielectric slit nonlocality would result in some minor changes of $\varepsilon_1$ and, as a consequence, some minor changes of *h*. However, the final equation (21) and the derivation presented in Appendix B will remain unchanged. One can see in Fig. 6 that the slit width *h* varies within about 1 nm range for all three dielectrics as the wavelength varies from 1535 nm to 1635 nm. As a practical implication, this can be used for fine tuning of the slit width by tuning the wavelength to achieve the matching condition set by Eq. (21). In reality, there will be some non-zero residual loss in the resonant slit waveguide due to a small impact of residual inter-band electron transitions and surface roughness. However, the impact of the inter-band electron transitions in the short wavelength infrared range is negligibly smaller relative the intra-band electron transitions while the surface roughness problem can be mitigated by using advanced deposition techniques such as atomic layer deposition. Thus, by quenching the energy dissipation due to the intra-band electron transitions in metals, one can dramatically reduce propagation losses for plasmons with deep sub-wavelength confinement. Although use of the slit waveguides for transporting plasmons is limited by its propagation distance within the metal film thickness, the slit waveguides with quenched energy dissipation can become a building block for a disruptive technology for truly nanometer-scale ultra-high quality factor plasmonic resonator cavities and electrically pumped ultra-low threshold and high slope-efficiency plasmonic lasers.

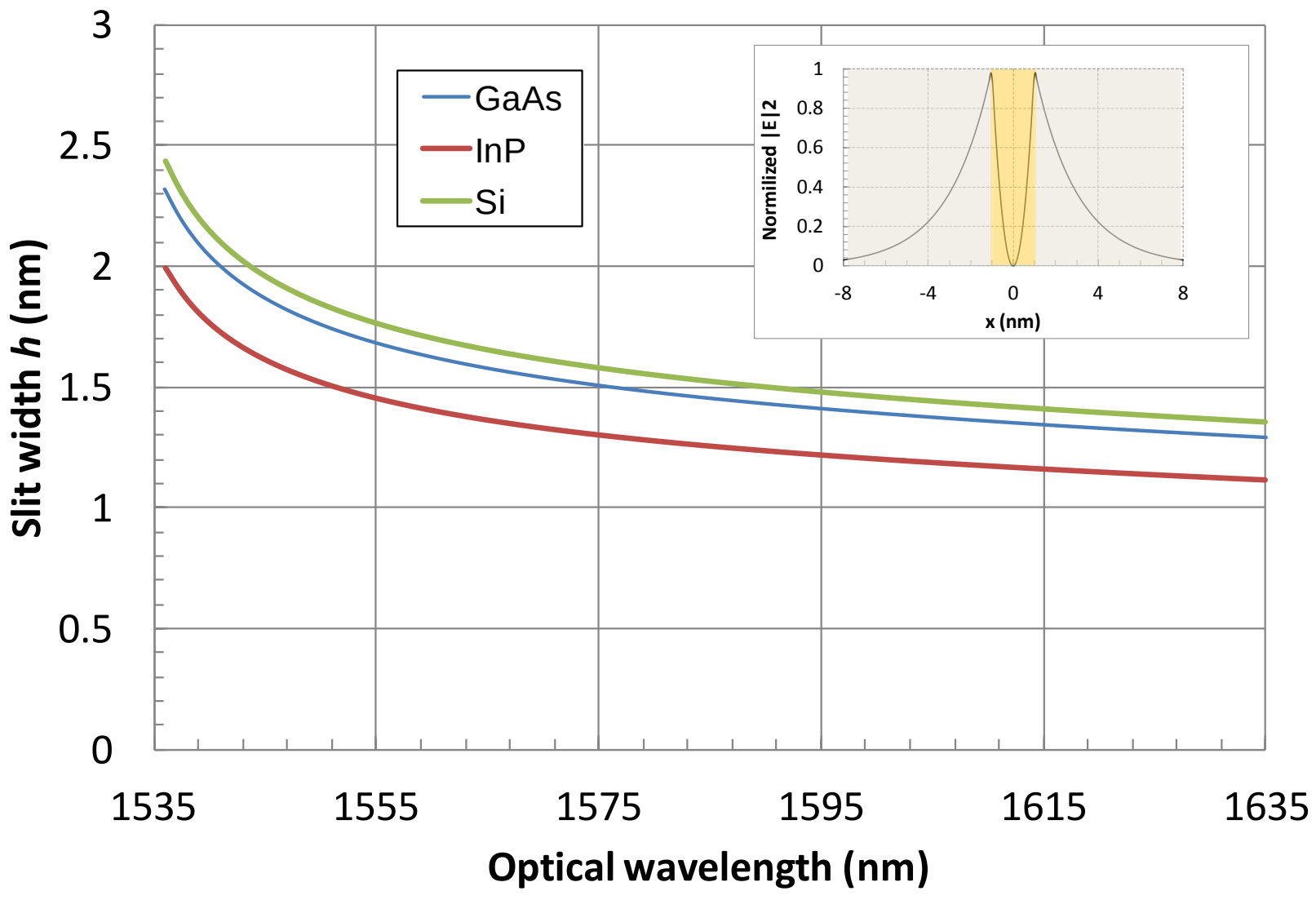

Fig. 6. Wavelength dependence of slit width *h*, according to Eq. (21), in the slit waveguide shown in Fig. 5 capable of supporting resonant plasmons with zero energy dissipation associated with intra band electron transitions. The slit waveguide comprises a *d* = 20 nm thick Au film with crystalline GaAs (blue), InP (red), and Si (green) slits. The inset shows an example of lossless plasmon mode profile for the slit waveguide with *h* = 2 nm Si slit width at 1543.7 nm wavelength with 20 nm thick Au film. The Au film Si slit areas are marked with grey and yellow colors respectively.

This is so because (unlike the existing plasmon lasers [5] – [7] where most of the pump power has to be spent just to overcome extremely high losses in metals, making their lasing efficiency poor and their pump threshold high) plasmonic lasers based on the resonant slit waveguides would allow us to eliminate the burden of extremely high losses, resulting in both dramatic reduction of lasing threshold and drastic increase of lasing efficiency. A straightforward way of building a cavity out of a resonant slit waveguide would be creating a Bragg grating in the slit with a small periodic variation of $\varepsilon_1$ along the *z*-axis with period of $\pi/b_0$.

## VII. Conclusion

It has been shown that there exists a special type of evanescent wave with resonant wave numbers capable of propagating in metal thin films with nonlocality in normal direction to film surface with zero energy dissipation associated with intra band electron transitions. Energy dissipation for these resonant evanescent waves is quenched due to destructive interference of spatially dispersed partial contributions to the electric displacement field in a nonlocal metal film over the film thickness. A dispersion relation coupling wave number parameters and dielectric response for a nonlocal thin film is derived, revealing wave number band gaps and band passes. Resonant wave numbers, for the evanescent waves with quenched energy dissipation, turn to be located at interfaces (edges) between a band gap and band passes in wave number space. It has been pointed out that resonant quenching of energy dissipation (RQED) can occur in films with thicknesses of less than the nonlocality scale length or more than, but still on the order of, the nonlocality scale length. RQED can no longer exist in metal films whose thickness exceeds a cutoff length or in metal films with local dielectric permittivity. It is demonstrated that the RQED effect can be used for designing a new type of plasmonic waveguides, such as a slit waveguide representing a metal film with a narrow slit filled with a dielectric, to achieve near zero propagation losses for a plasmon mode with few nanometer scale confinement. Such waveguides with quenched energy dissipation can become a building block for a disruptive technology in designing truly nanometer-scale: a) intra-chip interconnects with a length limited to the cutoff length, b) ultra-high quality factor plasmonic resonator cavities, and c) electrically pumped ultra-low threshold and high slope-efficiency plasmonic lasers.

## Appendix A: Dynamic equations for a TM wave in a metal film with nonlocality

Taking the curl of both of sides of the 1st equation (6a) and substituting the $\nabla\times\mathbf{H}(\mathbf{r},\omega)$ term from the 2nd equation (6a) into it, we obtain,

$$\nabla\times\nabla\times\mathbf{E}(\mathbf{r},\omega)-\mu\frac{\omega^2}{c^2}\mathbf{D}(\mathbf{r},\omega)=0. \tag{A1}$$

Using that $\nabla\times\nabla\times\mathbf{E}=\nabla(\nabla\cdot\mathbf{E})-\nabla^2\mathbf{E}$, one can write Eq. (A1) as

$$\nabla(\nabla\cdot\mathbf{E})-\nabla^2\mathbf{E}-\mu\frac{\omega^2}{c^2}\mathbf{D}(\mathbf{r},\omega)=0\,. \tag{A2}$$

On the other hand, from the 1st equation **1b** we have

$$\nabla\cdot\mathbf{D}=\nabla\cdot(\hat{\varepsilon}\mathbf{E})=(\nabla\hat{\varepsilon})\cdot\mathbf{E}+\hat{\varepsilon}(\nabla\cdot\mathbf{E})=0\,, \tag{A3}$$

where $\hat{\varepsilon}$ is the integral operator for the dielectric permittivity defined in Eq. (7). Introducing an inverse operator $\hat{\varepsilon}^{-1}$ (such that $\hat{\varepsilon}^{-1}\hat{\varepsilon}$ yields the identity operator) and taking $\hat{\varepsilon}^{-1}$ of both sides of Eq. (A3), we have $\nabla\cdot\mathbf{E}=-\hat{\varepsilon}^{-1}(\nabla\hat{\varepsilon})\cdot\mathbf{E}$. Substituting the $\nabla\cdot\mathbf{E}$ term from the later into Eq. (A2), one can write,

$$\nabla\left[\hat{\varepsilon}^{-1}(\nabla\hat{\varepsilon})\cdot\mathbf{E}\right]+\nabla^2\mathbf{E}+\mu\frac{\omega^2}{c^2}\mathbf{D}(\mathbf{r},\omega)=0\,. \tag{A4}$$

We will be looking for a solution of Eq. (A4) in a form of a transverse magnetic (TM) wave whose electric field vector is lying in the (*x*,*z*) plane,

$$\mathbf{E}(\mathbf{r},\omega)=\boldsymbol{\mathcal{E}}(z,\omega)\exp(ik_x x)\,, \tag{A5}$$

where $\boldsymbol{\mathcal{E}}=(\mathcal{E}_x,0,\mathcal{E}_z)$ and $k_x$ is an *x*-coordinate component of the wave vector $\mathbf{k}$. Note that the wave in Eq. (A5) can always be expanded in Fourier integral over the plane waves propagating in *z*-direction with continuously varying $k_z$, being the *z*- coordinate component of the wave vector $\mathbf{k} = (k_x, 0, k_z)$, but fixed $k_x$. Therefore, using Eq. (8), one can write the displacement vector induced by the $\boldsymbol{\mathcal{E}}(z,\omega)$ field as

$$\boldsymbol{\mathcal{D}}(z,\omega) = \int \varepsilon(\zeta,\omega)\boldsymbol{\mathcal{E}}(z-\zeta,\omega)d\zeta\,, \tag{A6}$$

where $\mathbf{D}(\mathrm{r},\omega) = \boldsymbol{\mathcal{D}}(z,\omega)\exp(ik_x x)$ and

$$\varepsilon(\zeta,\omega) = \frac{1}{2\pi}\int \varepsilon(k_x,k_z,\omega)\exp(-ik_z\zeta)dk_z \tag{A7}$$

with $\varepsilon(k_x,k_z,\omega)$ defined in Eq. (9). Using that now the integral operator $\hat{\varepsilon}$, defined by Eq. (A6), is independent of *x* and *y*, one can easily derive from Eq. (A4) an equation for $\mathrm{E}_z$ field component,

$$\frac{\partial}{\partial z}\left[\hat{\varepsilon}^{-1}\frac{\partial(\hat{\varepsilon}\boldsymbol{\mathcal{E}}_z)}{\partial z}\right] = \left(k_x^2 - \mu\frac{\omega^2}{c^2}\hat{\varepsilon}\right)\boldsymbol{\mathcal{E}}_z\,. \tag{A8}$$

Finally, re-writing Eq. (A8) as a set of two 1st order equations and using the explicit form of the operator $\hat{\varepsilon}$, we have

$$\begin{aligned}\frac{\partial D_z}{\partial z} &= \frac{\omega}{2\pi c}\int \varepsilon(\zeta,\omega)f(z-\zeta,\omega)d\zeta,\\ \frac{\partial f}{\partial z} &= \frac{2\pi c}{\omega}\left(k_x^2\boldsymbol{\mathcal{E}}_z - \mu\frac{\omega^2}{c^2}D_z\right),\end{aligned} \tag{A9}$$

where $D_z = \hat{\varepsilon}\boldsymbol{\mathcal{E}}_z = \int \varepsilon(\zeta,\omega)\boldsymbol{\mathcal{E}}_z(z-\zeta,\omega)d\zeta$ and $f = \frac{2\pi c}{\omega}\hat{\varepsilon}^{-1}\frac{\partial D_z}{\partial z}$. By using the residue theorem, integration over $k_z$ in Eq. (A7), with $\varepsilon(k_x,k_z,\omega) = \varepsilon_\infty + \varepsilon_{\mathrm{intra}}(k_x,k_z,\omega)$ and $\varepsilon_{\mathrm{intra}}(k_x,k_z,\omega)$ defined in Eqs. (9) and (10), yields the following expression for the nonlocal dielectric permittivity function $\varepsilon(\zeta,\omega)$,

$$\varepsilon(\zeta,\omega) = \varepsilon_\infty\delta(\zeta) + \frac{i\omega_p^2}{2a}\exp(ia|\zeta|), \tag{A10}$$

where $a^2 = \frac{1}{\beta^2}(\omega^2 + i\gamma\omega) - k_x^2$. The nonlocal dielectric permittivity function in Eq. (A10) has a clear physical meaning. It consists of the two parts. The first one is a local dielectric permittivity function part equal to $\varepsilon_\infty$ times the delta-function. The second one is a nonlocal dielectric permittivity function part with an exponentially decaying dielectric response having the nonlocality scale length (the distance $\zeta$ at which the nonlocal dielectric function part is reduced by a factor of $e$ relative to its maximum at $\zeta = 0$) of

$$r_{\mathrm{nl}} = [\mathrm{Im}(a)]^{-1} > 0\,. \tag{A11}$$

One can see that in the limit of the nonlocality parameter $\beta^2 \to 0$, the nonlocal dielectric function in Eq. (A7) reduces to a fully local dielectric function

$$\varepsilon(\zeta,\omega)=\left(\varepsilon_\infty-\frac{\omega_p^2}{\omega^2+i\gamma\omega}\right)\delta(\zeta),$$

in consistence with Eqs. (9) and (10) at $\beta^2=0$. Note that for a metal slab (or film) of limited thickness *d* lying parallel to the (*x*,*y*) plane, the integration with respect to $\zeta$ in Eqs. (A9) goes over the slab (or film) thickness *d*. That is,

$$D_z=\hat{\varepsilon}\mathcal{E}_z=\int_{-d/2}^{d/2}\varepsilon(\zeta,\omega)\mathcal{E}_z(z-\zeta,\omega)d\zeta. \tag{A12}$$

The validity of the integral form in Eq. (A12) stems from the first principles of the quantum mechanical representation of the dielectric permittivity function $\varepsilon(\mathbf{k},\omega)$ for a medium with limited dimensions [28]. Indeed, based on Eq. (12.6) in Ref. [28], for the case of isotropic medium, one can see that $\varepsilon(\mathbf{k},\omega)$ can be easily re-written as an integral (over the medium volume) of some function of space coordinate $\mathbf{r}$ times $\exp(-i\mathbf{kr})$. This function of space coordinate $\mathbf{r}$, being a Fourier image of $\varepsilon(\mathbf{k},\omega)$ divided by the medium volume is, consequently, by definition, the dielectric permittivity function in the spatial domain. For our case of $\mathbf{k}=(k_x,0,k_z)$ with fixed $k_x$, such integral representation of $\varepsilon(\mathbf{k},\omega)$ reduces to the following form,

$$\varepsilon(k_x,k_z,\omega)=\int_{-d/2}^{d/2}\varepsilon(\zeta,\omega)\exp(-ik_z\zeta)d\zeta. \tag{A13}$$

Multiplying Eq. (A13) by $\exp(ik_z z)$, using that $D_z(k_z,\omega)=\varepsilon(k_x,k_z,\omega)\mathcal{E}_z(k_z,\omega)$, where $D_z(k_z,\omega)$ and $\mathcal{E}_z(k_z,\omega)$ are the Fourier transforms of $D_z(z,\omega)=\sum_{k_z}D_z(k_z,\omega)\exp(ik_z z)$ and $\mathcal{E}_z(z,\omega)=\sum_{k_z}\mathcal{E}_z(k_z,\omega)\exp(ik_z z)$, and summing over $k_z$, we arrive to Eq. (A12). Thus, the dynamic equations (A9) for a TM wave in a metal slab (or film) of a limited thickness *d*, lying parallel to the (*x*,*y*) plane, read as follows,

$$\begin{aligned}\frac{\partial D_z}{\partial z}&=\frac{\omega}{2\pi c}\int_{-d/2}^{d/2}\varepsilon(\zeta,\omega)f(z-\zeta,\omega)d\zeta,\\ \frac{\partial f}{\partial z}&=\frac{2\pi c}{\omega}\left(k_x^2\mathcal{E}_z-\mu\frac{\omega^2}{c^2}A\right),\end{aligned} \tag{A14}$$

where $D_z=\hat{\varepsilon}\mathcal{E}_z=\int_{-d/2}^{d/2}\varepsilon(\zeta,\omega)\mathcal{E}_z(z-\zeta,\omega)d\zeta$. The $\mathrm{E}_x$ component of the TM wave is then found through the 2nd Eq. (6a) as

$$\hat{\varepsilon}\mathcal{E}_x(z) = \frac{i}{k_x}\frac{\partial D_z}{\partial z}. \qquad \text{(A15)}$$

**Appendix B: Mode equations and field profile for a TM-mode in a slit waveguide with nonlocality**

Let us assume that a *z*-component of the electric field $\mathbf{E}(\mathbf{r},\omega) = (E_x, 0, E_z)$ for a TM plasmon wave propagating in the metal films on the right and left sides from the dielectric core (slit) in the slit waveguide shown in Fig. 5 has a form,

$$E_z(x,z,\omega) = \begin{cases} C_3(\omega)\exp[ik_{x0}(x - h/2) + b_0 z], & x \geq h/2, \\ -C_3(\omega)\exp[-ik_{x0}(x + h/2) + b_0 z], & x \leq -h/2, \end{cases} \qquad \text{(B1)}$$

where $C_3(\omega)$ is defined in Eq. (14), $k_{x0}$ and $b_0$ are the resonant wave numbers (described in Section V) satisfying the dispersion relation in Eq. (19) in which $k_x = k_{x0}$, $b = b_0$ and $\varepsilon_B = \varepsilon_{B0}$, to yield zero energy dissipation ($\mathrm{Im}(\varepsilon_{B0}) = 0$). *h* is the slit width. In the waveguide dielectric core (slit), the wave has a form,

$$E_z(x,z,\omega) = A(\omega)[\exp(ik_{x1}x) - \exp(-ik_{x1}x)]\exp(ib_0 z), \quad |x| < h/2, \qquad \text{(B2)}$$

where $A(\omega)$ and $k_{x1}$ are the amplitude and wave number *x*-component in the core. Let us also assume that *y*-component of the magnetic field $\mathbf{H}(\mathbf{r},\omega)$ for the TM plasmon wave outside the waveguide core has a form

$$H_y(x,z,\omega) = \begin{cases} B(\omega)\exp[ik_{x0}(x - h/2) + b_0 z], & x \geq h/2, \\ B(\omega)\exp[-ik_{x0}(x + h/2) + b_0 z], & x \leq -h/2, \end{cases} \qquad \text{(B3)}$$

whereas in the dielectric core, it is

$$H_y(x,z,\omega) = H(\omega)[\exp(ik_{x1}x) + \exp(-ik_{x1}x)]\exp(ib_0 z), \quad |x| < h/2, \qquad \text{(B4)}$$

where $B(\omega)$ and $H(\omega)$ are some amplitudes in the metal film and dielectric core (slit) respectively. The $E_z(x,z,\omega)$ and $H_y(x,z,\omega)$ fields are the two tangential components that must be continuous through the $x = \pm h/2$ interfaces. Therefore, from Eqs. (B1) and (B2) it has to be,

$$C_3(\omega) = A(\omega)[\exp(ik_{x1}h/2) - \exp(-ik_{x1}h/2)]. \qquad \text{(B5)}$$

Likewise, from Eqs. (B3) and (B4) it has to be,

$$B(\omega) = H(\omega)[\exp(ik_{x1}h/2) + \exp(-ik_{x1}h/2)]. \qquad \text{(B6)}$$

For a TM wave with $\mathbf{E}(\mathbf{r},\omega)=(E_x,0,E_z)$ and $\mathbf{H}(\mathbf{r},\omega)=(0,H_y,0)$, the $B(\omega)$ and $C_3(\omega)$ amplitudes are coupled to each other through an equation

$$\frac{\partial H_y(x,z,\omega)}{\partial x}=i\frac{\omega}{c}\mathrm{D}_z(x,z,\omega), \tag{B7}$$

which directly follows from the 2nd equation (6a), where $\mathrm{D}_z(x,z,\omega)$ is a $z$-component of $\mathbf{D}(\mathbf{r},\omega)$. Next, we substitute Eq. (B3) into the left-hand side of Eq. (B7), use that $\mathrm{D}_z(x,z,\omega)=\varepsilon_{B0}E_z(x,z,\omega)$ in the metal films on the right and left sides of the slit core (with $\varepsilon_{B0}$ being the resonant dielectric response of the metal film described in Section V), and substitute $E_z(x,z,\omega)$ from Eq. (B1) into the right-hand side of Eq. (B7). We then obtain that Eq. (B7) is satisfied both at $x\geq h/2$ and $x\leq -h/2$ when

$$B(\omega)=\frac{\omega\varepsilon_{B0}}{ck_{x0}}C_3(\omega). \tag{B8}$$

Similarly, for the dielectric slit core ($|x|<h/2$), we substitute Eq. (B4) into the left-hand side of Eq. (B7), use that $\mathrm{D}_z(x,z,\omega)=\varepsilon_1 E_z(x,z,\omega)$ in the slit core (where $\varepsilon_1$ is the dielectric permittivity of the dielectric slit core), and substitute $E_z(x,z,\omega)$ from Eq. (B2) into the right-hand side of Eq. (B7). We the obtain,

$$H(\omega)=\frac{\omega\varepsilon_1}{ck_{x1}}A(\omega). \tag{B9}$$

Substitution of Eqs. (B8) and (B9) into Eq. (B6) yields

$$C_3(\omega)=\frac{\varepsilon_1 k_{x0}}{\varepsilon_{B0}k_{x1}}A(\omega)\left[\exp(ik_{x1}h/2)+\exp(-ik_{x1}h/2)\right]. \tag{B10}$$

Finally, compatibility of Eq. (B5) and Eq. (B10) requires that

$$\left[\exp(ik_{x1}h/2)-\exp(-ik_{x1}h/2)\right]=\frac{\varepsilon_1 k_{x0}}{\varepsilon_{B0}k_{x1}}\left[\exp(ik_{x1}h/2)+\exp(-ik_{x1}h/2)\right]. \tag{B11}$$

Re-writing Eq. (B11), one has

$$\exp(ik_{x1}h)=\frac{\varepsilon_{B0}k_{x1}+\varepsilon_1 k_{x0}}{\varepsilon_{B0}k_{x1}-\varepsilon_1 k_{x0}}, \tag{B12}$$

where $k_{x1}$ satisfies the dispersion relation in the dielectric slit core,

$$k_{x1}^2+b_0^2=\frac{\omega^2}{c^2}\varepsilon_1\mu, \tag{B13a}$$

whereas the resonant wave numbers $k_{x0}$ and $b_0$ are coupled by the dispersion relation in Eq. (19), i.e.,

$$k_{x0}^2 + b_0^2 = \frac{\omega^2}{c^2}\varepsilon_{B0}\mu. \tag{B13b}$$

Equations (B12) and (B13) represent a set of mode equations for a TM-mode in the slit waveguide shown in Fig. 5. Thus, for a given resonant evanescent wave (with the resonant wave numbers $k_{x0}$, $b_0$ and the resonant dielectric response $\varepsilon_{B0}$), there exists a resonant width $h$ for the dielectric slit, defined by Eq. (B12), that supports a TM-mode with quenched energy dissipation. The solution for the $E_x$ component is found through Eq. (18) (Section IV) using the 1st Eq. (14) with $C_1 = \varepsilon_{B0}C_3$, which yields

$$E_x(x,z,\omega) = \begin{cases} -\dfrac{b_0}{k_{x0}}E_z(x,z,\omega), & |x| \geq \dfrac{h}{2}, \\ -\dfrac{b_0}{k_{x1}}E_z(x,z,\omega), & |x| < \dfrac{h}{2}. \end{cases} \tag{B14}$$

For a negative and real $\varepsilon_{B0}$, positive and real $\varepsilon_1$, positive imaginary $k_{x0}$ and imaginary $k_{x1}$ of both signs Eq. (B12) has always a non-trivial solution for $h$ (for large enough $b_0^2 > \varepsilon_1\mu\omega^2/c^2$). Thus, from Eqs. (B1), (B2), (B5) and (B14) the field $\mathbf{E}(\mathbf{r},\omega) = (E_x, 0, E_z)$ profile for the TM plasmonic mode with zero energy dissipation (associated with the intra-band electron transitions) in the resonant slit waveguide shown in Fig. 5 is as follows,

$$E_z(x,z,\omega) = \begin{cases} -C_3\exp\left[|k_{x0}|(x+h/2) + ib_0z\right], & x \leq -h/2, \\ -2A(\omega)\sinh(k_{x1}^I x)\exp(ib_0z), & |x| < h/2, \\ C_3\exp\left[-|k_{x0}|(x-h/2) + ib_0z\right], & x \geq h/2, \end{cases} \tag{B15}$$

where $C_3 = -2A(\omega)\sinh(k_{x1}^I h/2)$ and $k_{x1}^I = \mathrm{Im}(k_{x1})$, whereas $E_x(x,z,\omega)$ is expressed through $E_z(x,z,\omega)$ according to Eq. (B14).

## References

[1] S. A. Maier, *Plasmonics: Fundamentals and Applications*, (Springer Science+Buisiness Media LLC, 2007), Chap. 2.

[2] D. Sarid, "Long-range surface-plasma waves on very thin metal films," *Phys. Rev. Lett.,* **47**(26), pp 1927-1930 (1981).

[3] P. Markoš, I. Rousochatzakis, C. M. Soukoulis, "Transmission losses in left-handed materials," *Phys. Rev. E,* **66**(045601), pp 1-4 (2002).

[4] Th. Koschny, J. Zhou, C. M. Soukoulis, "Magnetic Response and Negative Refractive Index of metamaterials'" *Proc. SPIE,* **6581**(658103), pp 1-15 (2007).

[5] M. A. Noginov et al, "Enhancement of surface plasmons in an Ag aggregate by optical gain in a dielectric medium," *Opt. Lett.* **31**(20), pp 3022-3024 (2006).

[6] M. A. Noginov, "Compensation of surface plasmon loss by gain in dielectric medium," *J. Nanophotonics,* **2**(021855), pp 1-17 (2008).

[7] R. F. Oulton et al, "Plasmon lasers at deep subwavelength scale," *Nature,* **461**(08364), pp 629-632 (2009).

[8] L. Thylén, P. Holmström, A. Bratkovsky, J. Li, S.-Y. Wang, "Limits on Integration as Determined by Power Dissipation and Signal-to-Noise Ratio in Loss-Compensated Photonic Integrated Circuits Based on Metal/Quantum-Dot Materials," *IEEE J. Quantum Electron.,* **46**(4), pp 518-524 (2010).

[9] J. B. Khurgin, G. Sun, "In search of the illusive lossless metal," *Appl. Phys. Lett.*, **96**(181102), pp 1-3 (2010).

[10] V. L. Bonch-Bruevich, S. G. Kalashnikov, *Physics of Semiconductor*, (VEB Deuthscher Verlag der Wissen-Schaften, Berlin, 1982), Chap. XVIII, Paragraph 5.

[11] T.-C. Chiang, "Photoemission studies of quantum well states in thin films," *Surf. Sci. Rep.,* **39**(2000), pp 181-235 (2000).

[12] J. J. Paggel, T. Miller, T.-C. Chiang, "Quantum-well states as Fabry-Pérot modes in a thin-film electron interferometers," *Science,* **283**(12 March), pp 1709-1711 (1999).

[13] N. V. Smith, "Phase analysis of image states and surface states associated with nearly-free-electron band gaps," *Phys. Rev. B,* **32**(6), pp 3549-3555 (1985). See Eq.(4).

[14] L. Rettig, P. S. Kirchmann, U. Bovensiepen, "Ultrafast dynamics of occupied quantum well states in Pb/Si (111)," *New Journal of Physics*, **14**(023047), pp 1-17 (2012).

[15] L. Allen, J. H. Eberley, *Optical Resonance and Two-Level Atoms*, (Wiley-Interscience Publication, John Wiley and Sons, New-York – London – Sydney – Toronto, 1975), Chap. 1, Paragraph 3.

[16] J. M. McMahon, S. K. Gray, G. C. Schatz, "Nonlocal optical response of metal nanostructures with arbitrary shape," *Phys. Rev. Lett.*, **103**(097403), pp 1-4 (2009).

[17] J. M. McMahon, S. K. Gray, G. C. Schatz, "Calculating nonlocal optical properties of structures with arbitrary shape," *Phys. Rev. B*, **82**(035423), pp 1-12 (2010).

[18] P. B. Johnson, R. W. Christy, "Optical constants of the noble metals," *Phys. Rev. B,* **6**(12), pp 4370-4379 (1972).

[19] H. Ehrenreich, H. R. Philipp, "Optical properties of Ag and Cu," *Phys. Rev.,* **128**(4), pp 1622-1629 (1962).

[20] R. Cooper, H. Ehrenreich, H. R. Philipp, "Optical properties of noble metals. II", *Phys. Rev.*, **138**(2A), pp 494–507 (1965).

[21] A. L. Fetter, "Electrodynamics of a layered electron gas. I. Single layer," *Ann. Phys.,* **81**(2), pp 367-393 (1973). See page 392.

[22] E. A. Coronado, G. C. Schatz, "Surface Plasmon broadening for arbitrary shape nanoparticles: A geometric probability approach," *J. Chem. Phys.*, **119**(7), pp 3926-3934 (2003). Paragraph III.

[23] M. Liu, P. Guyot-Sionnest, "Synthesis and optical characterization of Au/Ag core/shell nanorods," *J. Phys. Chem.,* **108**(19), pp 5882-5888 (2004).

[24] W. E. Jones, K. L. Kleiwer, R. Fuchs, "Nonlocal theory of the optical properties of thin metallic films," *Phys. Rev.* **178**(3), pp 1201-1203 (1969).

[25] I. Lindau, P. O. Nilsson, "Experimental evidence of excitation for longitudinal plasmons by photons," *Phys. Letts.,* **31A**(7), pp 352-353 (1970).

[26] M. Bass, *Handbook of Optics, 3rd edition Vol. 4*, (McGraw-Hill, 2009).

[27] M. Bass, *Handbook of Optics, 2nd edition Vol. 2*, (McGraw-Hill, 1994).

[28] V. M. Agranovich, V. L. Ginzburg, *Spatial Dispersion in Crystal Optics and the Theory of Excitons*, (Interscience Publishers, a division of John Wiley & Sons, London - New York – Sydney, 1966), Chap. 4.